\documentclass[draftclsnofoot,onecolumn]{IEEEtran}

\usepackage{color,array,amsthm}
\usepackage{graphicx}
\usepackage{cite}
\usepackage{amsmath,amssymb,amsfonts}
\usepackage{algorithmic}
\usepackage{textcomp}
\usepackage{booktabs}
\usepackage{makecell}
\usepackage{caption}
\usepackage{url}
\usepackage{subcaption}

\begin{document}

\title{Optimal DC-link Voltage from Weight and Loss Perspective for eVTOLs} 

\author{Abhijit Kulkarni, Torbjörn Thiringer, and Remus Teodorescu
\thanks{This work was supported in part by the the OPENSRUM grant (223286)
from MSCA-IF European Commission and  Smart Battery Villum Investigator Grant (222860) from VILLUM FONDEN.}
\thanks{A. Kulkarni and R. Teodorescu are with the Department of Energy, Aalborg University, Aalborg, Denmark.}
\thanks{T. Thiringer is with the Department of Electrical Engineering, Chalmers University of Technology, Gothenburg, Sweden. }

}

%\author{Abhijit Kulkarni}
%\member{Member, IEEE}
%\affil{Department of Energy, Aalborg University, Denmark.} 

% \author{Torbjörn Thiringer}
% %\member{Fellow, IEEE}
% %affil{Chalmers University of Technology, Gothenburg, Sweden.} 

% \author{Remus Teodorescu}
% %\member{Fellow, IEEE}
% %\affil{Department of Energy, Aalborg University, Denmark.}

\markboth{}{}
\maketitle

\begin{abstract}
Electric vertical takeoff and landing (eVTOL) aircraft are emerging as a modern transportation solution aimed at reducing urban traffic congestion and improving the carbon footprint. The power architecture in eVTOLs is defined by the dc bus formed by the battery packs and the power converter used to drive eVTOL motors. A high dc bus voltage is preferred for the power architecture since it can reduce the weight of power cables for a given power rating. However, the impact of high dc bus voltage on the efficiency of the drivetrain power converter must be considered, since reduced efficiency leads to poor battery pack utilization. In this paper, a systematic optimization study is performed considering SiC-based inverter for the drivetrain power converter. Optimal value of dc bus voltage is determined considering the flight profile of eVTOLs. A power converter topology is proposed that can provide optimal performance and enhance the lifetime of the batteries along with providing better monitoring, diagnostics and protection. The optimization strategy is validated experimentally, demonstrating the proposed power architecture's ability to maximize efficiency while enhancing the safety of the battery energy storage system in eVTOLs.
\end{abstract}


\section{INTRODUCTION}
Electric mobility is undergoing a significant transformation, with growing numbers of electric passenger vehicles (EVs), buses, heavy-duty trucks, aircraft, and even ships. Advancements in lithium-ion battery technology, featuring increased energy densities and sophisticated battery management systems (BMS), have improved both performance and safety, driving this expansion \cite{lib_advances1,lib_advances2,lib_enabled_ev}. Within e-mobility, electric vertical takeoff and landing aircraft (eVTOLs) have received positive public reception. A survey from the European Union Aviation Safety Agency (EASA) \cite{easa_uam_2021} shows that a significant majority (83\%) of respondents were positively inclined towards the introduction of urban air mobility (UAM) solutions comprising of eVTOLs, commercial unmanned aerial vehicles (UAVs), etc., in their cities. Numerous aerospace companies are developing eVTOLs of different size and payload capacities \cite{evtol_comparison,flyingcars_uiuc}. Commercial operation of eVTOLs is expected as early as 2025.

\begin{figure}[htbp]
    \centering
    \includegraphics[width=0.8\linewidth]{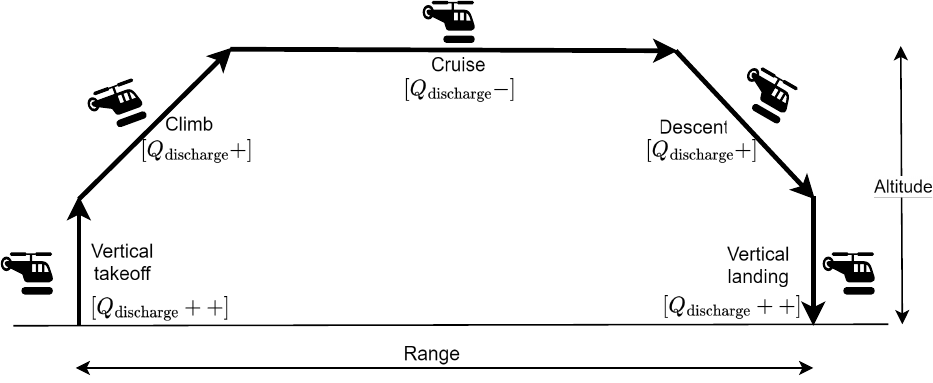}
    \caption{Typical mission profile of an eVTOL.}
    \label{fig:evtol_mission}
\end{figure}

EVTOLs are designed to revolutionize urban transportation by providing a sustainable and efficient solution to traffic congestion and carbon emissions. Applications envisioned for eVTOLs include air-taxis, air-ambulances, cargo delivery, etc. There is a considerable variety in the structure and payload carrying capacity of the eVTOLs \cite{evtol_comparison,flyingcars_uiuc}. For example, Lilium jet has a range of 203km, uses a battery pack rated for 38kWh and can carry upto 5 passengers \cite{evtol_comparison}. In contrast, Ehang 216 has a range of 35km, uses a battery pack of 17kWh and can accommodate upto 2 passengers \cite{ehang_ref_mdpi}. Despite these variations, eVTOLs share a common flight profile, as shown in Figure \ref{fig:evtol_mission}. This profile consists of five segments: vertical takeoff, climb, cruise, descent, and vertical landing \cite{flyingcars_uiuc,evtol_good_joule,evtol_kolar}. The discharge of Li-ion cells during each segment is also shown in Fig. \ref{fig:evtol_mission} using the parameter $Q_{discharge}$.
The power requirement is also very high during takeoff and landing segments while it is at a minimum during cruise. As indicated in a recent battery test dataset \cite{evtol_dataset}, the discharge power ratio between takeoff/landing and cruise can be as high as 3.3.

\begin{figure}[htbp]
    \centering
    \includegraphics[width=0.85\linewidth]{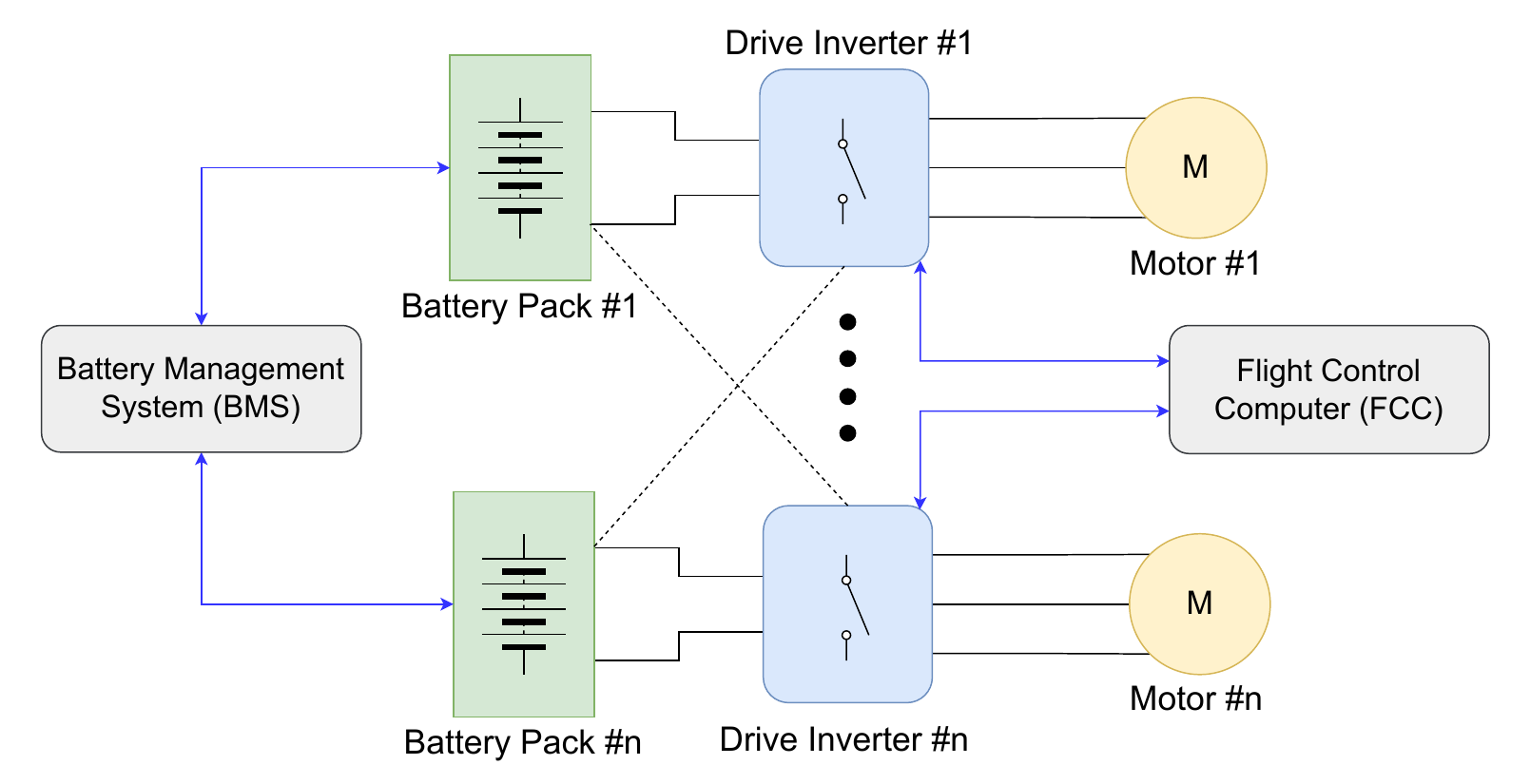}
    \caption{Conventional power architecture in a multi-rotor eVTOL with FCC and BMS.}
    \label{fig:power_arch_conv}
\end{figure}

The power architecture in eVTOLs mirrors the general structure found in EVs. However, as eVTOLs are typically multirotor systems, the battery pack is often divided into multiple units distributed throughout the aircraft. Each battery pack can power one or more motors in the multirotor configuration using either single-stage inverters or dual-stage power converters, which consist of a DC-DC converter followed by a DC-AC inverter \cite{inverter_ref,dcdc_dcac_fuelcell_evtol,doppler2024requirements,dc_ac_evtol_glasgow, power_electronics_evtol_good,evtol_kolar,sae_good_topology_battery}. Fig. \ref{fig:power_arch_conv} illustrates the power architecture in a multirotor eVTOL with distributed battery packs and inverters. The dotted lines represent potential interconnections between battery packs and drive inverters to ensure fail-safe operation. Notably, in some cases, a DC-DC converter is employed as an intermediate stage between the battery packs and the drive inverter. 
%The location of charging power circuit is not shown in  Fig. \ref{fig:power_arch_conv}. Also, for the sake of brevity, mechanical components such as gearbox that are associated with the motor-rotor are not shown.

Furthermore, in Fig. \ref{fig:power_arch_conv}, the battery management system (BMS) and flight control computer (FCC) are shown along with their connections to the drivetrain components. The BMS senses cell voltages, temperature, and pack current to estimate key states such as battery state of charge (SoC) and state of health (SoH). It also performs protective actions in safety-critical events like thermal runaway. The FCC, on the other hand, provides torque and speed command references to the drive inverter's controller, allowing for the extraction of desired power from the motors. These references vary depending on the segment of the mission profile shown in Fig. \ref{fig:evtol_mission}. It is important to note that the BMS and FCC typically operate independently \cite{jobypatent}.

The drive motors and inverters are typically located close together, while the battery packs may be situated at a distance. This necessitates the use of long interconnect power cables to connect them. To reduce cable weight, a high DC bus voltage is preferred \cite{power_electronics_evtol_good}. For a given power requirement, a higher DC voltage naturally reduces the current rating and, consequently, the cross-sectional area of the power cables. However, the impact of high DC bus voltage on drive inverter performance and overall system weight has not been systematically studied in the literature. As reported in \cite{power_electronics_evtol_good,evtol_kolar,sae_good_topology_battery}, the dc bus voltage for UAM applications is limited up to 800V. 

Consider the efficiency of the drive inverter, which is expected to be very high to ensure better utilization of battery energy. However, a high DC bus voltage may affect power converter efficiency due to its impact on switching and conduction losses. Any negative impact on efficiency due to high DC bus voltage is undesirable. Thus, an optimization study of the impact of increasing DC bus voltage at the system level is crucial. The impact of higher DC bus voltage on insulation requirements and the potential increase in cable weight due to thicker insulation should also be considered.

Since the mission profile demands varying power from the battery packs depending on the flight segment, a variable DC bus architecture may be preferable. This architecture allows the DC bus voltage to be reconfigured to not only provide the necessary power but also ensure maximum efficiency for the drivetrain power converters. Therefore, this paper proposes a  power architecture for eVTOLs, including reconfigurable smart battery packs and close interaction between the BMS and FCC to optimize battery pack performance, lifetime, and system safety. The optimization analysis is done considering that the drivetrain power converter is based on SiC devices. This is because wide bangap devices such as SiC MOSFETs have lower losses compared to traditional Si-based IGBTs for high voltage applications. Additionally, wide bandgap-based power devices are considered to have an overall superior performance especially in aerospace applications \cite{power_electronics_evtol_good, wbg_aerospace}. 
The specific contributions of this work can be summarized as follows:
\begin{itemize}
\item Selection of an optimal DC bus voltage beyond the current state-of-the-art by considering the interaction between DC bus voltage, power cable weight, and drive inverter efficiency.
\item A  power architecture designed to optimize battery pack performance and lifetime, taking into account the typical eVTOL mission profile.
\item Experimental validation using an SiC-based three-phase inverter.
\end{itemize}
This paper is organized as follows: Section II introduces a systematic optimization method for selecting the DC bus voltage in a wide bandgap-based drive inverter, aimed at optimizing both efficiency and cable weight. In Section III, a power architecture for eVTOLs is proposed that makes use of  the eVTOL mission profile to maximize drive inverter efficiency while enhancing diagnostics, safety, and the cycle life of battery packs. Reconfigurable battery packs are used in the proposed architecture. Section IV presents experimental results that validate the proposed DC bus selection approach. Finally, Section V discusses the conclusions and challenges associated with the proposed approach.

\section{DC Bus Voltage Optimization}
In this section, the impact of higher dc bus voltage on the efficiency of a single-stage drive inverter is analyzed first. The inverter topology considered is the standard two-level three-phase inverter as considered in earlier works \cite{inverter_ref,doppler2024requirements,dc_ac_evtol_glasgow, power_electronics_evtol_good,sae_good_topology_battery}. Then, the improvement in the cable weight due to increasing dc bus voltage is quantified. An objective function for optimization is developed considering the quantitative impacts of power losses and the cable weight, which is then solved to obtain an optimal dc bus voltage for the eVTOL drive train.

\subsection{Power loss model for SiC-based drive inverter}
\begin{figure}
    \centering
    \includegraphics[width=0.8\linewidth]{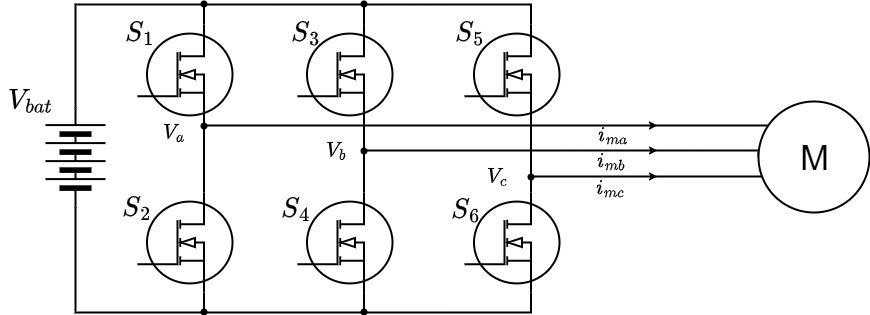}
    \caption{Conventional two-level three-phase  drive inverter feeding eVTOL motor.}
    \label{fig:inv3ph}
\end{figure}

Consider an SiC-based two-level three-phase inverter as shown in Fig. \ref{fig:inv3ph} driving an eVTOL motor. The power losses are mainly in the SiC devices and include conduction and switching losses.
To compute the conduction losses, the rms current of the switches is to be computed. For switching losses, the average current and the dc bus voltage will have a direct impact. Since, the inverters typically supply the motors without any additional inductors, there are no other losses apart from the losses in the dc bus capacitors. 

Assuming that the motor currents are sinusoidal, they are expressed as follows
\begin{align}
    i_{ma} &= I_m\sin(\omega t) \label{eq:ima} \\
    i_{mb} &= I_m\sin\left(\omega t - \frac{2\pi}{3}\right) \label{eq:imb} \\
    i_{mc} &= I_m\sin\left(\omega t + \frac{2\pi}{3}\right) \label{eq:imc} 
\end{align}
In (\ref{eq:ima}) -- (\ref{eq:imc}), motor current amplitude is represented by $I_m$ and the fundamental frequency corresponding to the rotor speed is represented by $\omega$. 
Assuming sine-triangle modulation, the duty ratio references are provided in (\ref{eq:da}) -- (\ref{eq:db}), where $m$ represents the modulation index ($0<m\leq 1$). The phase angle $\phi$ is included to account for the phase difference between motor currents and the applied inverter voltage. Note that for other modulation methods such as space vector pulse-width modulation (SVPWM), there will be an additional common mode term that can be ignored for the differential-mode analysis. Thus, the rms and average currents that will be calculated in this section will practically apply to both sine-triangle PWM and SVPWM.
\begin{align}
    d_{a} &= 0.5 + 0.5m\sin(\omega t + \phi) \label{eq:da} \\
    d_{b} &= 0.5 + 0.5m\sin\left(\omega t - \frac{2\pi}{3} + \phi\right ) \label{eq:db} \\
    d_{c} &= 0.5 + 0.5m\sin\left(\omega t + \frac{2\pi}{3} + \phi\right) \label{eq:dc} 
\end{align}
Due to the symmetricity of the currents and modulation reference signals, all the switches will have the same average and rms currents. The deadtime effect is ignored in this analysis to result in closed-form expressions that will be used for developing the objective function for the optimization method. The duty ratio expressions contain a phase angle $\phi$ indicative of the motor power factor.

Let the switches in Fig. \ref{fig:inv3ph} have a switching function each that defines whether a switch is on or off. This is defined for the first switch $S_1$ as follows,
\begin{equation}
\tilde{S}_1 = 
\begin{cases} 
1 & \text{when } S_1 \text{ is ON} \label{eq:sw_fn_s1}\\
0 & \text{when } S_1 \text{ is OFF} 
\end{cases}
\end{equation}
If $T_s$ is the switching period, the rms current for a single switching period can be written as,
\begin{align}
    i_{rms,Ts}^2 &= \frac{1}{T_s}\int_0^{T_s} i_{ma}^2(t) \tilde{S}_1(t) dt \label{eq:irmsts1}\\
     &=  i_{ma}^2(t)d_a(t) \label{eq:irmsts2}
\end{align}
Eq. (\ref{eq:irmsts1}) is simplified to (\ref{eq:irmsts2}) considering that the motor current does not change much during a switching period. Additionally, the average of the switching function $\tilde{S}_1$ is the duty ratio $d_a$ from the definition. 

The rms current across the full fundamental cycle for the switch $S_1$ is determined using the following relation,

\begin{align}
    i_{rms}^2 &= \frac{1}{T_0}\int_0^{T_0} i_{ma}^2(t)d_a(t) dt \notag \\
    &= \frac{1}{T_0}\int_0^{T_0} I_m^2\sin^2(\omega t)[0.5 + 0.5m\sin(\omega t + \phi)] \notag \\
    &= \frac{1}{T_0}\int_0^{T_0}  0.5I_m^2\sin^2(\omega t) \notag \\
    &~~~~+ \frac{1}{T_0}\int_0^{T_0}  0.5mI_m^2\sin^2(\omega t) \sin(\omega t + \phi) \label{eq:irmst0_1}
\end{align}
It can be proved that the second term in (\ref{eq:irmst0_1}) equals zero and only the first term contributes towards the rms current. It can be further simplified to result in the final expression,
\begin{equation}
    i_{rms} = \frac{I_m}{2} \label{eq:irms_final}
\end{equation}
The rms current of every switch in Fig. \ref{fig:inv3ph} is given by (\ref{eq:irms_final}), which can be seen to be independent of the dc bus voltage.

Now, to compute the switching losses, average current is needed. Average switch current across a single switching period is given by
\begin{align}
    i_{avg,Ts}^2 &= \frac{1}{T_s}\int_0^{T_s} i_{ma}(t) \tilde{S}_1(t) dt \label{eq:iavgts1}\\
     &=  i_{ma}(t)d_a(t) \label{eq:iavgts2}
\end{align}
As done in case of rms current, (\ref{eq:iavgts2}) is used to compute the average switch current for the fundamental period using
\begin{align}
    i_{avg} &= \frac{1}{T_0}\int_0^{T_0} i_{ma}(t)d_a(t) dt \notag \\
    &= \frac{1}{T_0}\int_0^{T_0} I_m\sin(\omega t)[0.5 + 0.5m\sin(\omega t + \phi)] \notag \\
    &= \frac{1}{T_0}\int_0^{T_0}  0.5I_m\sin(\omega t) \notag \\
    &~~~~+ \frac{1}{T_0}\int_0^{T_0}  0.5mI_m\sin(\omega t) \sin(\omega t + \phi) \label{eq:iavgt0_1}
\end{align}
It can be proved that the first term in (\ref{eq:iavgt0_1}) equals zero and only the second term contributes to the average current. It can be further simplified to result in the final expression,
\begin{equation}
    i_{avg} = \frac{mI_m\cos\phi}{4} \label{eq:iavg_final}
\end{equation}
Due to symmetry, all the switches will have the same average curent as in (\ref{eq:iavg_final}).

Now, the expression for conduction loss is given by
\begin{equation}
    P_{cond} = \frac{I_m^2}{4}r_{ds,on} \label{eq:pcond}
\end{equation}
The on-resistance of the SiC MOSFET $r_{ds,on}$ and the rms currents are used to arrive at the expression for the conduction loss in (\ref{eq:pcond}).

The switching loss for the SiC MOSFETs depends on the energy loss during turn-on and turn-off transitions. These energies, namely, $E_{on}$ and $E_{off}$ are specified in the device datasheet at specified test dc bus voltage and switching currents. The energies need to be scaled to the actual dc bus voltage \cite{chalmers_sic_loss} and average switch current to result in an average switching loss given by
\begin{equation}
    P_{sw} = f_{sw}(E_{on} + E_{off})\left(\frac{V_{dc}}{V_{ref}}\right)^{k_v}\left(\frac{mI_m\cos\phi}{4}\right) \label{eq:psw}
\end{equation}
Thus, the total power loss as a function of the dc bus voltage is found as
\begin{align}
    &P_{loss}(V_{dc}) = 
    \frac{I_m^2}{4}r_{ds,on} + f_{sw}(E_{on} + E_{off})\left(\frac{V_{dc}}{V_{ref}}\right)^{k_v}\left(\frac{mI_m\cos\phi}{4}\right) \label{eq:ptot}
\end{align}
It can be observed that with the increasing $V_{dc}$, the power losses may increase due to the exponential term $k_v$ in (\ref{eq:ptot}), even though the motor current may decrease for a given power rating. It also depends on if the SiC device has predominantly conduction losses or switching losses. 

\subsection{Modeling the impact of cable weight}
\begin{figure}[htbp]
    \centering
    \includegraphics[width=0.65\linewidth]{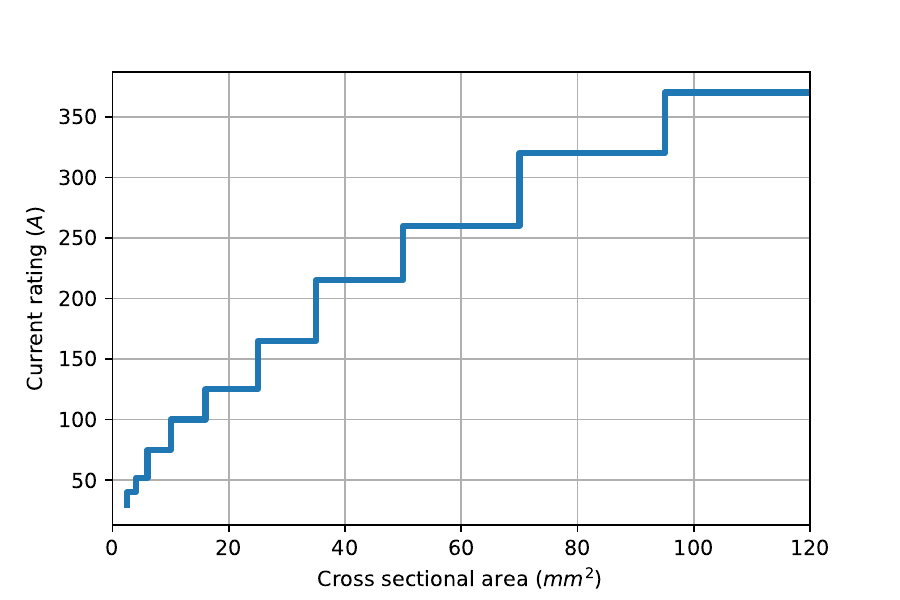}
    \caption{Cable cross sectional area versus current rating for 1.5kV rated dc power cables.}
    \label{fig:cable_current}
\end{figure}

% Points to cover in this part
% \begin{itemize}
%     \item Explain the table but a reference is a must
%     \item Add an equation about battery current and emphasize that we are actually interested in dc current because that cable is the longest. Motor and inverter cable are short.
%     \item Indicate that insulation grade does not change from 900V to say 2kV.
%     \item Show the graph for 40kW or whatever power rating
% \end{itemize}
The current rating of the power cables is defined based on the cross sectional area. Fig \ref{fig:cable_current} shows the current rating of power cables as a function of their cross-sectional area. Note that the current ratings appear in a stepped manner due to the discrete cross sectional areas available in commercial cables. The current ratings in  Fig \ref{fig:cable_current} are based on the 1.5kV dc cables with a derating factor of about 75\% than what is mentioned in \cite{current_rating_ref}.
The cable weight from the battery packs to the drive inverter plays a critical role in the overall weight compared to the cables from the drive inverter to the motor. This is because typically the drive inverter and motor are located close to each other and as a result the cable lengths are small. Now, using the above curve, it is possible to determine the copper wire radius as the dc bus voltage is varied for a given power rating.

\begin{figure}[htbp]
    \centering
    \includegraphics[width=0.65\linewidth]{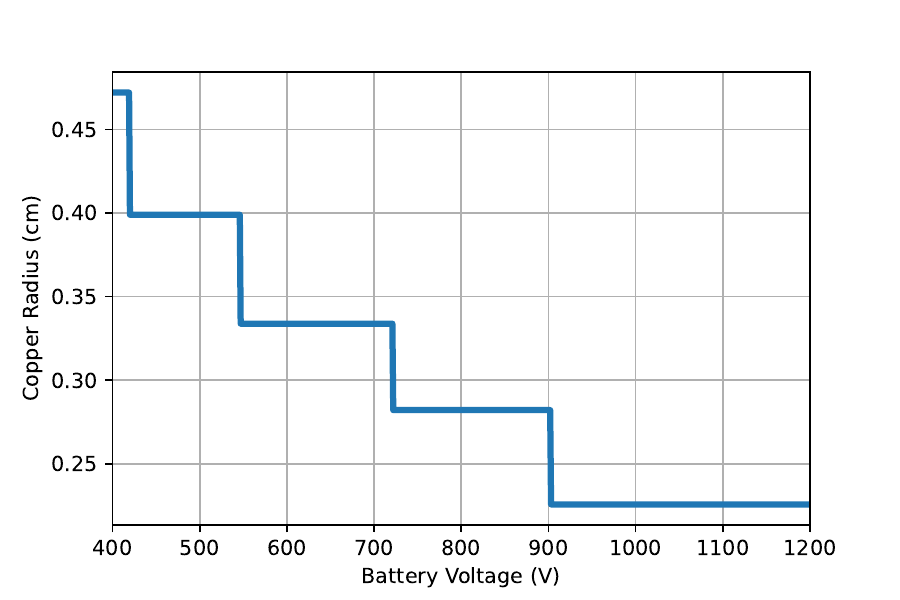}
    \caption{Radius of the copper in the power cable versus dc bus (battery) voltage for a 121hp motor.}
    \label{fig:cu_radius_vs_vdc}
\end{figure}

Consider that the eVTOL motor is rated for 121hp as an example \cite{bldc_motor_data}. This corresponds to an output power requirement of 89 kW. If the dc bus voltage is varied from 400V to 1500V, the impact on the copper diameter used in the power cable is illustrated in Fig. \ref{fig:cu_radius_vs_vdc}.

As the voltage rating is increased, the insulation thickness may increase for the power cables. As it can be seen in standards such as IEC 60502-1 \cite{iec60502}, typically the insulation thickness changes at discrete levels in commercial power cables. Thus, upto $1.2~kV$ range, the insulation thickness will be at a certain level which will increase when the voltage range goes beyond $1.2~kV$. However, for higher current rating, the insulation thickness is relatively higher compared to lower current rated cables  \cite{iec60502}. Thus, the impact of higher voltage rating on insulation thickness may be offset by the reduced current rating.
Nevertheless, it is important to quantify the theoretical increase in the insulation thickness to understand if its effect on cable weight change can be ignored as the dc bus voltage is varied.

In \cite{insulation}, an analytical method is developed to compute the insulation thickness for a given applied voltage and a given insulator material. The mathematical expression proposed in \cite{insulation} is given by
\begin{equation}
t_i = r_c \left( \exp\left(\frac{K V_{\text{max}} t_v}{\alpha r_c}\right) - 1 \right) + C \label{eq:insulation}
\end{equation}
In (\ref{eq:insulation}), $r_c$ is the radius of the copper in the power cable, $C$ is a constant given by $C = 0.1~cm$ for applied voltages of $<20~kV$, $V_{max}$ is the applied voltage and in this paper it is from $400~V$ to $1200~V$. The parameter $t_v$ is the void within the insulator that results in a weakest point causing breakdown and hence affects the value of the insulation thickness. It is taken as $t_v = 50~\mu m$ along with $\alpha = 0.340~kV$ as per \cite{insulation}. Finally,   $K$ is defined as the coefficient expressing the field enhancement within the cavity. Its value is determined
by the shape of the cavity. For example, for spherical voids, 
\begin{equation}
    K = \frac{3\epsilon _r}{1+2\epsilon _r} \label{eq:K_val}
\end{equation}
In (\ref{eq:K_val}), $\epsilon _r$ is the relative permittivity of the insulation material. 

\begin{figure}[htbp]
    \centering
    \includegraphics[width=0.65\linewidth]{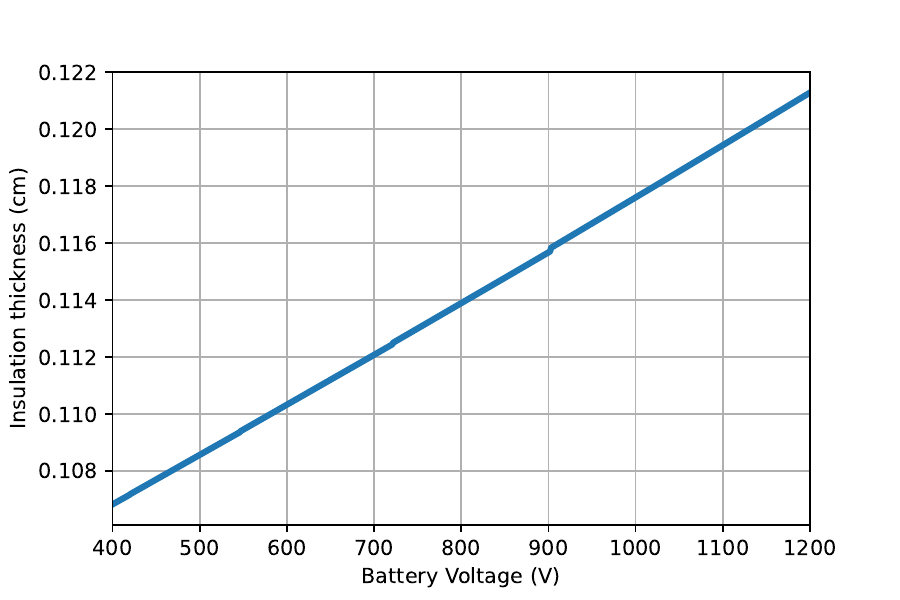}
    \caption{Power cable insulation thickness versus dc bus (battery) voltage.}
    \label{fig:insulation_vs_vdc}
\end{figure}

Using this analytical model, the insulation thickness is plotted as the dc bus voltage increases. It is shown in Fig. \ref{fig:insulation_vs_vdc}.
It can be observed from \ref{fig:insulation_vs_vdc} that an increase in the dc bus voltage from $500~V$ to $1~kV$ will increase the theoretical insulation thickness by 10\%. On the other hand, the same amount of increase in dc bus voltage, results in a reduction of the copper radius by a factor of  43\% and a reduction in copper volume by a factor of 68\%.

\subsection{Proposed optimization}
Using the power loss model and the quantitative impact of the dc bus voltage on the cable weight, an objective function is developed to choose an optimum value of the dc bus voltage. 
Cross sectional area of the copper cable ($A_{cu}$) is proportional to the cable weight. Thus, it is used as a parameter is forming the objective function for the optimization. 

The objective function can be setup as
\begin{equation}
\begin{aligned}
    &\text{Minimize} \quad f(V_{dc}) = \beta A_{cu} + (1-\beta) P_{loss} \\
    &\text{subject to:} \\
    &\quad \phantom{xxxxxxx}0 < \beta < 1 
    %&\quad \phantom{xxxxxxx}V_{dc} \leq 1500\text{V}
\end{aligned}
\label{eq:objfn}
\end{equation}

If the weighting factor $\beta$ is chosen to be small, higher priority is given to minimizing the power loss. Conversely, a larger value for $\beta$ results in a $V_{dc}$ that prioritizes minimizing the cable weight. Any aircraft has a known parameter called power-to-weight ratio, expressed in $kW/kg$, which indicates the amount of engine or motor power necessary to carry a certain payload. Multirotor eVTOLs generally have a higher power-to-weight ratio compared to fixed-wing aircraft, reflecting their greater sensitivity to weight. Thus, in this work, it is preferred to generate an optimum solution for a higher value of $\beta$ or at least $\beta > 0.5$. 
%The dc bus voltage is limited to $1.5~kV$ since higher voltages would lead to the selection of Si IGBTs that have lower efficiency than SiC MOSFETs. 

\subsection{Results}
The battery pack voltage is considered to vary from $400~V$ to $1500~V$. Since a single SiC device will not be optimal in this entire voltage range, the optimization study on the impact of the dc bus voltage variation is performed using three SiC MOSFETs. For a dc bus voltage ($V_{dc}$) of  $V_{dc}\leq500~V$, power losses are determined using the device E4M0025075J2 \cite{sic750v} rated for $750~V$. For the dc bus voltage in the range $500~V<V_{dc}<1000~V$, power losses are determined using the device AIMZHN120R010 \cite{sic1200v} rated for $1200~V$. Finally, for $V_{dc}>1000~V$ and upto $1500~V$, power losses are determined using the device C2M0045170P \cite{sic1700v} rated for $1700~V$. Key parameters of the three SiC MOSFETs are summarized in Table \ref{tab:SiCMOSFET} using their respective datasheets. 
\begin{table}[ht]
\caption{SiC MOSFET Characteristics}
\centering
\begin{footnotesize}
\begin{tabular}{ccccc}
\toprule
\textbf{SiC MOSFET} & \textbf{$V_{DSS}$} & \textbf{$I_{D}$} & \textbf{$R_{DS(on)}$} & \textbf{$E_{oss}$}  \\
\midrule 
E4M0025075J2 \cite{sic750v} & 750 V & 84 A & 25 $m\Omega$ & 23 $\mu J$  \\
\midrule 
AIMZHN120R010 \cite{sic1200v} & 1.2 kV & 202 A & 8.7 $m\Omega$ & 107 $\mu J$  \\
\midrule 
C2M0045170P \cite{sic1700v} & 1.7 kV & 75 A & 40 $m\Omega$ & 139 $\mu J$   \\
\bottomrule
\end{tabular}
\end{footnotesize}
\label{tab:SiCMOSFET}
\end{table}
For computing the power loss as derived in (\ref{eq:ptot}), the motor currents, switching frequency and the device parameters from the datasheets are needed. An illustrative example is shown next considering a sample eVTOL motor from \cite{evtol_motor_spec} with slight modifications in the ratings. The key parameters of the motor considered in this paper listed below in Table \ref{tab:motor_param}.

\begin{table}[htbp]
\centering
\caption{Motor Parameters used for $V_{dc}$ optimization.}
\begin{tabular}{ll}
\toprule
\textbf{Parameter} & \textbf{Value} \\
\midrule
$P_{motor}$ & 57.6 kW \\
Rotor speed ($\omega$) & 328.6 rad/s \\
$K_t$ & 0.6 Nm/A \\
$\eta _{motor}$ & 91.8\% \\
$I_m$ & 194.6 A \\
\bottomrule
\end{tabular}
\label{tab:motor_param}
\end{table}

Now, with the semiconductor device parameters in Table \ref{tab:SiCMOSFET} along with the motor and device currents, the variation of the objective function in (\ref{eq:objfn}) versus dc bus voltage $V_{dc}$ is determined for different values of the weight $\beta$. This is done as explained below.
\begin{itemize}
    \item Motor torque and speed are assumed to remain constant irrespective of the DC bus voltage.
    \item Motor currents are computed using the motor torque and the torque constant ($K_t$).
    \item The fundamental voltage required for the motor is determined using the motor speed and the back EMF constant ($K_e = K_t$).
    \item The effective modulation index ($m$) is computed for a given DC bus voltage ($V_{dc}$) using the motor's fundamental voltage and the PWM      method.
    \item The SiC MOSFET RMS and average currents are determined using the modulation index and motor currents.
    \item The DC bus current is calculated for a given mechanical output power and DC bus voltage, assuming motor and inverter efficiencies.  The copper area $A_{cu}$ is then computed using the dc bus current and Fig. \ref{fig:cable_current}.
    \item These parameters ($A_{cu}$, device currents, switching frequency) are used in (\ref{eq:ptot}) and (\ref{eq:objfn}) along with the MOSFET datasheet parameters to determine the value of the objective function for each $V_{dc}$.
\end{itemize}

\begin{figure}[htbp]
    \centering
    \includegraphics[width=0.65\linewidth]{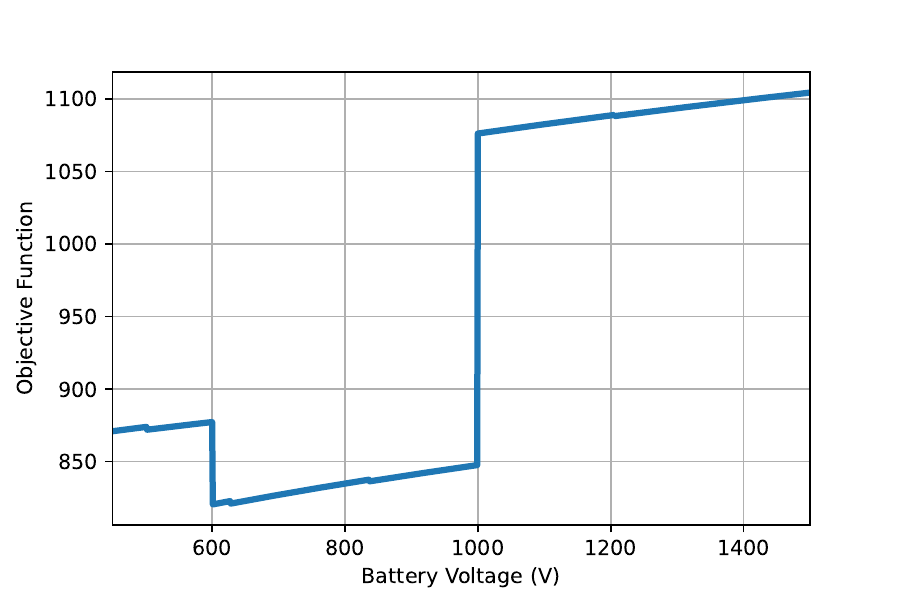}
    \caption{Variation of the objective function versus dc bus voltage when $\beta = 0.2$.}
    \label{fig:fobj_vs_vdc_0.2}
\end{figure}

Fig. \ref{fig:fobj_vs_vdc_0.2} shows the variation of the objective function as the dc bus or the battery pack voltage is varied from $450~V$ to $1500~V$. This is when $\beta = 0.2$. For this case, higher weight is given to the efficiency of the drive inverter.
It can be observed that the optimum is when $V_{dc}\approx600~V$. The discontinuities in the objective function seen in Fig. \ref{fig:fobj_vs_vdc_0.2} are due to the change in the device parameters since three SiC devices are used within the range of dc bus voltage variation.

\begin{figure}[htbp]
    \centering
    \includegraphics[width=0.65\linewidth]{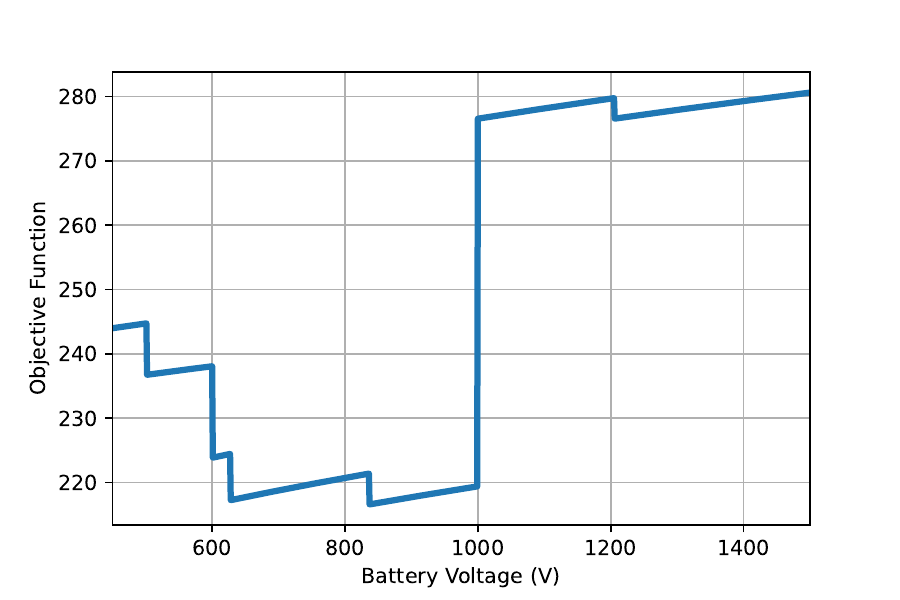}
    \caption{Variation of the objective function versus dc bus voltage $\beta = 0.8$.}
    \label{fig:fobj_vs_vdc_0.8}
\end{figure}

\begin{figure}[htbp]
    \centering
    \includegraphics[width=0.65\linewidth]{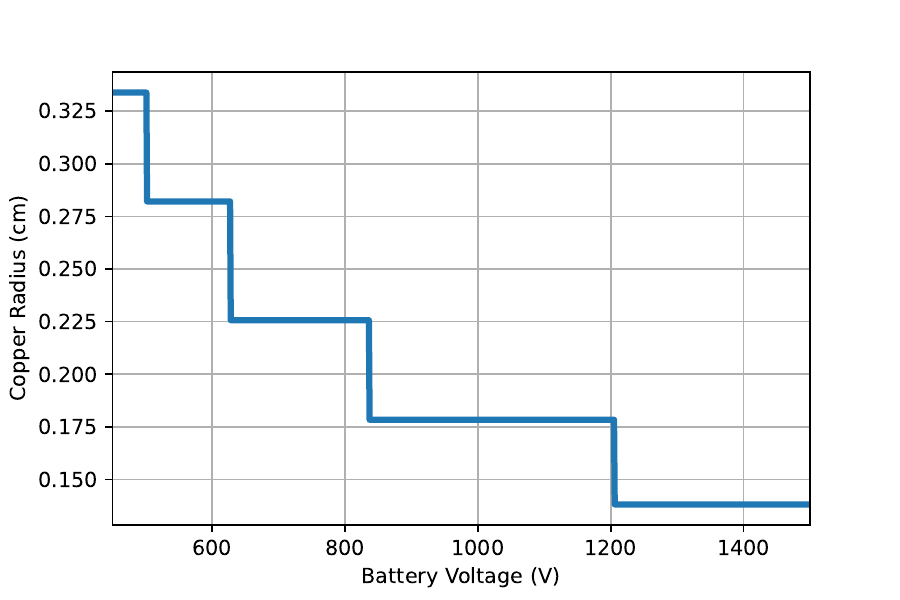}
    \caption{Variation of the radius of the power cable versus dc bus voltage for $\beta = 0.8$.}
    \label{fig:cu_radius_vs_vdc_optim}
\end{figure}

\begin{figure}[htbp]
    \centering
    \includegraphics[width=0.65\linewidth]{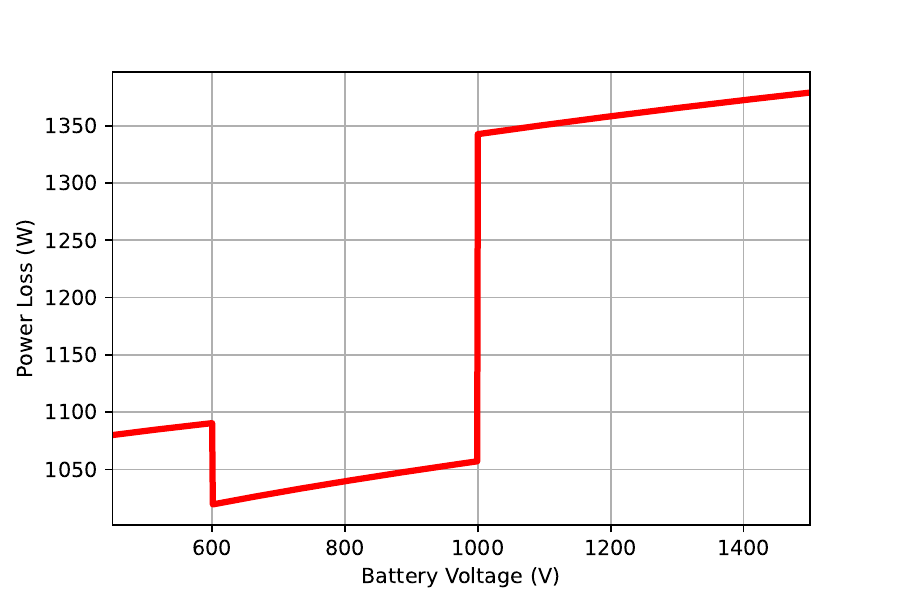}
    \caption{Variation of the inverter power loss versus dc bus voltage for $\beta = 0.8$.}
    \label{fig:ploss_vs_vdc_optim}
\end{figure}

When there is a higher priority given to the cable weight by increasing $\beta = 0.8$, the objective functions displays a different variation with respect to $V_{dc}$. This is shown in Fig. \ref{fig:fobj_vs_vdc_0.8}.
The optimum $V_{dc}$ for this case is $V_{dc}\approx840~V$. Note that the value of the objective function is also considerably lower compared to the result in Fig. \ref{fig:fobj_vs_vdc_0.2} indicating that the improvement in the weight is considerably higher. The copper cable radius for the given power and voltage variations are shown in Fig. \ref{fig:cu_radius_vs_vdc_optim}. 
Since the increase in the value of the objective function between $V_{dc}=840~V$ and $V_{dc}=1000~V$, it is proposed to use an optimal dc bus voltage for the eVTOL as $V_{dc,opt}=1000~V$. With this the cable weight reduces one step further than for the case with $V_{dc}=840~V$, which is at the marginal level. Thus, it can be observed that an optimal dc bus voltage can be derived once the motor parameters are known. The total power loss in the semiconductor devices versus dc bus voltage is also shown in   Fig. \ref{fig:ploss_vs_vdc_optim}.
The summary of the impact of different weights $\beta$ to the objective function is provided in Table \ref{tab:optim_summary}.
\begin{table}[ht]
\caption{Summary of dc bus voltage optimization}
\centering
\begin{footnotesize}
\begin{tabular}{ccccc}
\toprule
\textbf{$\beta$} & \textbf{$V_{dc,opt}~(V)$} & \textbf{$P_{loss}~(W)$} & \textbf{$r_{c}~(cm)$} & \textbf{$a_{w}~(cm^2)$}  \\
\midrule 
0.2  & 600 & 1019 & 0.282 & 0.25  \\
\midrule 
0.8  & 1000 & 1056 & 0.178 & 0.1  \\
\bottomrule
\end{tabular}
\end{footnotesize}
\label{tab:optim_summary}
\end{table}
As it can be observed, with the dc bus voltage increased to $1~kV$, the power losses are increased only by $37~W$, while the cross sectional area is reduced by a factor of 2.5. Since the cable cross sectional area is proportional to the cable weight, the weight of the cables is reduced by the same factor of 2.5. The proposed approach can be done for any eVTOL system once the motors are designed. The optimal dc bus voltage can then be used to configure the batteries into modules and packs.

% next subsection - optimization results.
% Motor rms, peak current calculations and modulation index calculation.
% Optimization result at different beta values.
% next section - overall power architecture
% explain with the drawing and advanced BMS features.
% final section - experiments.

\section{Proposed Power Architecture}
\begin{figure}[htbp]
    \centering
    \includegraphics[width=0.65\linewidth]{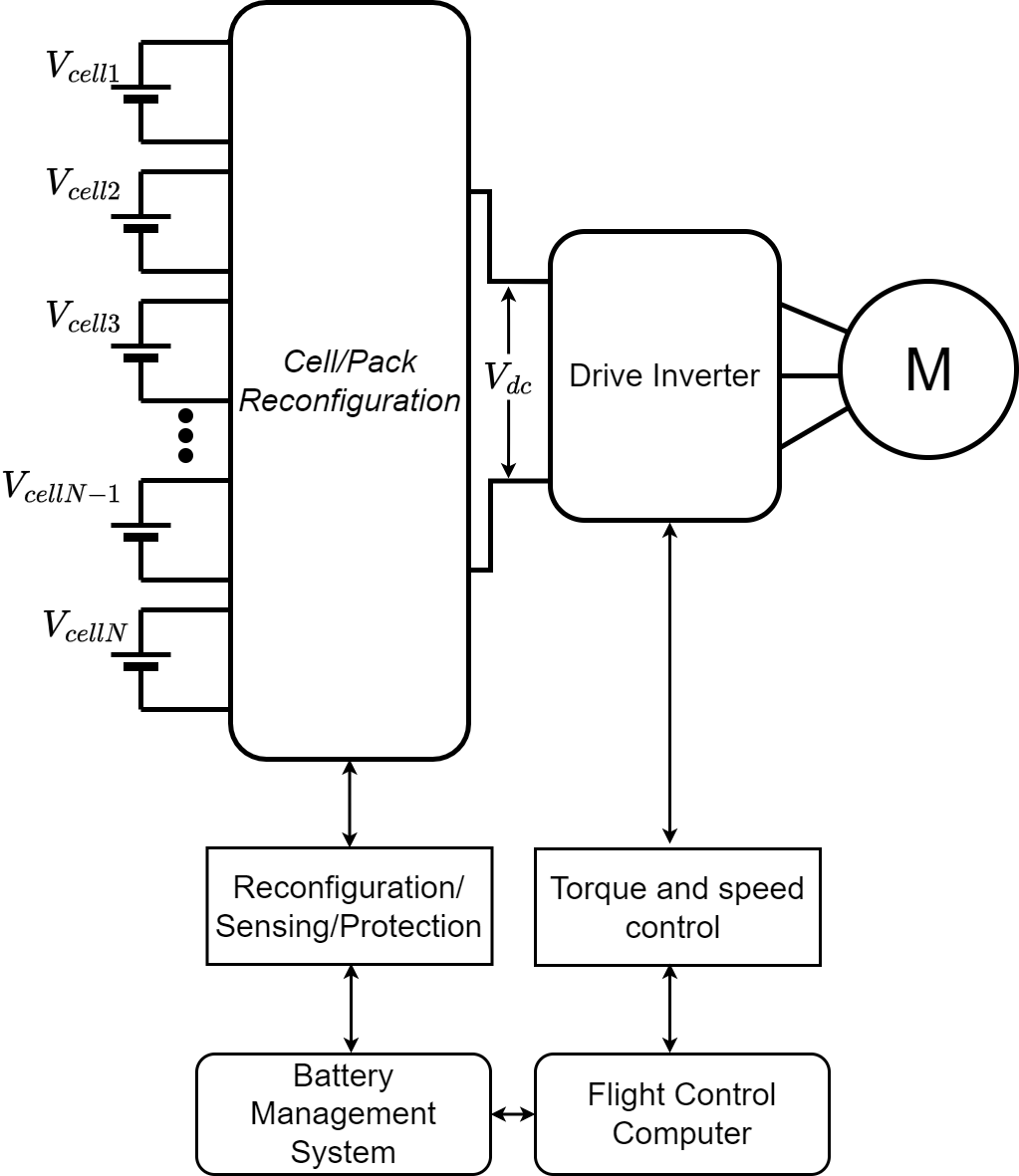}
    \caption{Proposed power architecture for an eVTOL.}
    \label{fig:proposed_arch}
\end{figure}

As discussed in Section II, a high $V_{dc}$ is preferable to ensure reduced eVTOL power cable weight. EVTOLs have the well-defined mission profile as shown in Fig. \ref{fig:evtol_mission}, wherein highest current is drawn from the battery packs only during the hovering phases of vertical takeoff and landing. Thus, during the other operating modes of climb, cruise and descent, the power requirement is lower in terms of both torque and speed. This brings out an important question about controlling the battery pack voltage during different flight modes. If it is possible to control the dc bus voltage, it would help in enhancing the power converter efficiency further. For example, consider the longest operating phase in an eVTOL, which is the cruise phase. If the dc bus voltage is reduced, the battery packs can still support the necessary torque and speed while reducing the power losses. As seen in Fig. \ref{fig:ploss_vs_vdc_optim}, a reduction in voltage will reduce the switching losses and hence improve the efficiency. This will enhance the battery utilization and prolong the SoC drop. Thus, in this work, a reconfigurable dc bus architecture is proposed where the battery cells are electronically reconfigured to provide the highest voltage during takeoff and landing while providing an optimal lower voltage during the other flight modes. This has the benefit of improving the system efficiency while ensuring that the cable weight is low due to high dc bus voltage during flight phases with high power demand.
A conceptual illustration of the proposed dc bus architecture is shown in Fig. \ref{fig:proposed_arch}.
It is shown for one motor in a multirotor eVTOL architecture. It can be compared with the conventional architecture shown in Fig. \ref{fig:power_arch_conv}. The distinguishing feature of the proposed architecture is the reconfigurable cell/pack block connected across the battery cells. It can also be observed that now there is a closed-loop interaction between the BMS and FCC unlike the conventional architecture. 

\begin{figure}[htbp]
    \centering
    \includegraphics[width=0.3\linewidth]{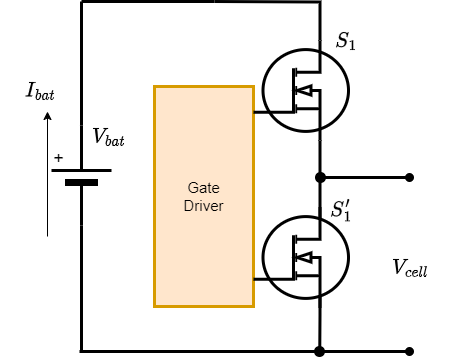}
    \caption{Cell reconfiguration using a half bridge.}
    \label{fig:halfbridge}
\end{figure}

Electronic reconfiguration can be done at the cell level or at a module level where multiple cells are connected in series/parallel combination. This is different from using a dc-dc converter since reconfiguration uses only semiconductor devices to either insert a cell or group of cells into the pack or to bypass them. Filters and specific PWM methods are not necessary. The most basic reconfiguration at cell level is shown in Fig. \ref{fig:halfbridge} where a half-bride topology is used to either insert a cell or bypass it from the battery pack.

\begin{figure}[htbp]
    \centering
    \includegraphics[width=.7\linewidth]{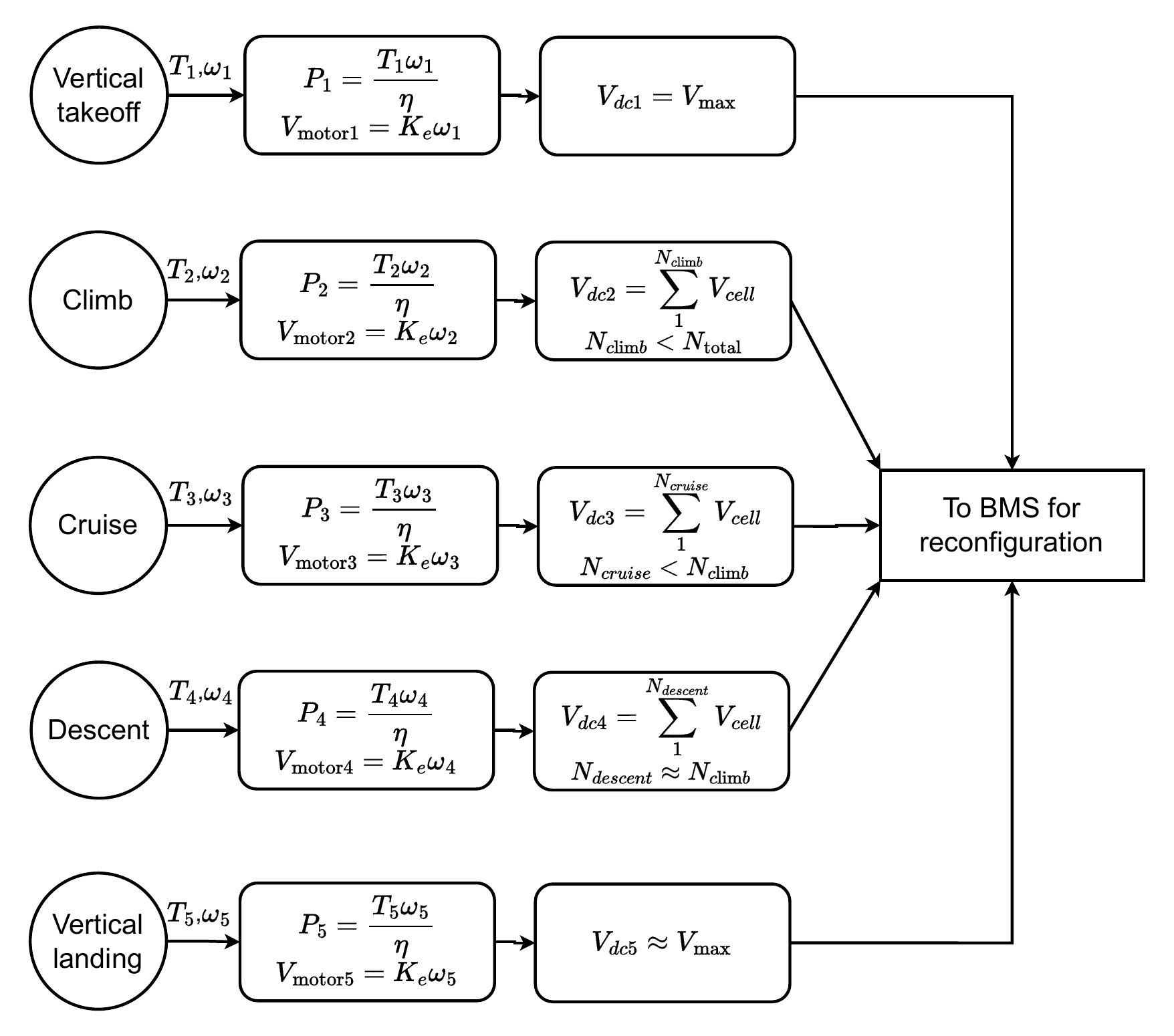}
    \caption{Control of the dc bus voltage during different phases of an eVTOL.}
    \label{fig:reconfig_flowchart}
\end{figure}

As can be seen from Fig. \ref{fig:halfbridge}, when the top switch $S_1$ is ON, $V_{cell}=V_{bat}$ whereas when the bottom switch $S_1'$ is ON, $V_{cell}\approx 0$ and both the switches are complementary. There are earlier works that have discussed the use of reconfigurable cells/modules \cite{reconfig_changfu, reconfig_efficiency,iecon_ak,reconfig_changfu2}. Reconfigurable topologies are quite suitable for aerospace application such as eVTOL due to their mission profile as discussed in this paper. The reconfiguration can be implemented as illustrated in the conceptual diagram in Fig. \ref{fig:reconfig_flowchart}. 
Here, for each phase of the flight mission profile the  required torque ($T$) and speed ($\omega$) are used to compute the power required and the motor phase voltage required. The dc bus voltage is proportional to the motor phase voltage and hence it can be used to compute the number of reconfigurable cells/modules that can be connected in series. Due to the mission profile, the number of cells needed is maximum for takeoff and landing phases, while it is minimum for the cruise phase. This is indicated in Fig. \ref{fig:reconfig_flowchart}. This also indicates the close interaction between the FCC and BMS since the torque and speed values are determined by the FCC, which are then used to obtain the optimal number of cells to be used for each phase of the flight.

%explain figure
%show reconfiguration
%subsection on advantages of the proposed approach with a flowchart.

\section{Experimental Results}
\begin{figure}
    \centering
    \includegraphics[width=1\linewidth]{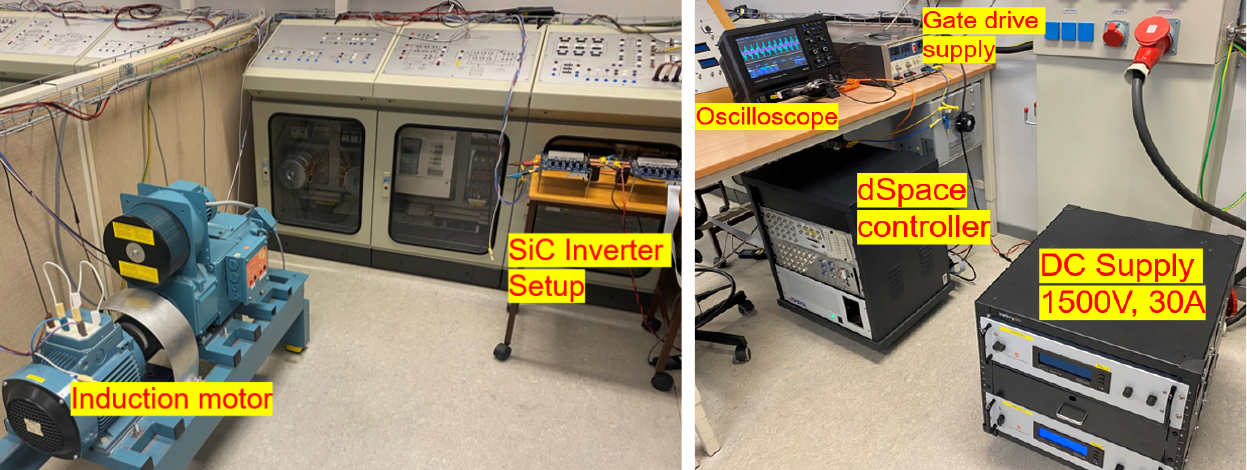}
    \caption{Full experimental setup showing the induction motor, dc supply, dSpace controller and measuring equipment.}
    \label{fig:fullsetup}
\end{figure}

In this section, the power loss model developed in Section II is validated considering real SiC devices. The optimization discussed in Section II, relies on the power loss model and the cable weight model to arrive at an optimal dc bus voltage that is used to select the active cells in the proposed architecture in Fig. \ref{fig:proposed_arch}. Therefore, the accuracy of the optimization method will be high if the power loss model is correct. 
\begin{figure}[htbp]
    \centering
    \includegraphics[width=0.75\linewidth]{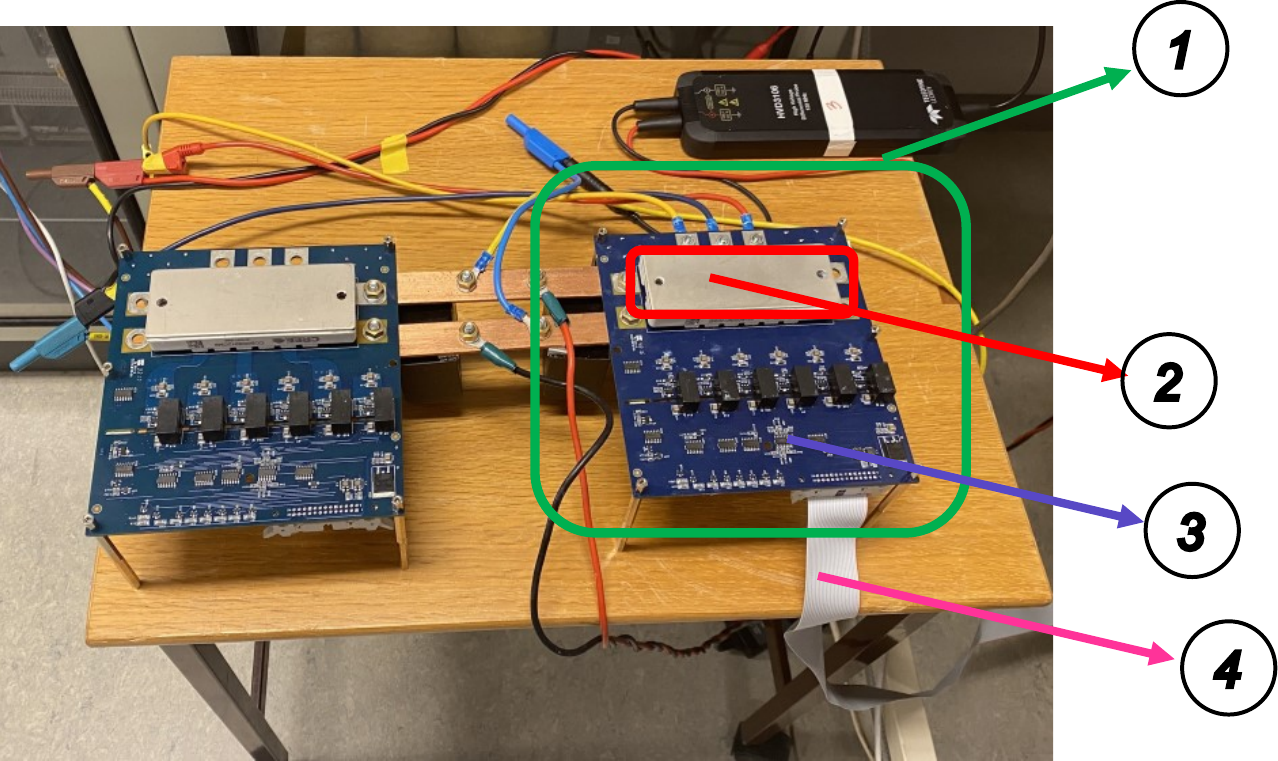}
    \caption{SiC based inverter used for experimental validation. (1) one three-phase inverter, (2) SiC power module, (3) gate driver, and (4) ribbon cable for the controller.}
    \label{fig:SiCinverter}
\end{figure}
The three-phase inverter used for experimental validation uses a $1200~V$, $50~A$ SiC module CCS050M12CM2 \cite{datasheet_sic}.  Fig. \ref{fig:SiCinverter} shows the inverter PCB used for the experimental validation. The gate driver used is CGD15FB45P1 from Wolfspeed \cite{datasheet_gd}. The inverter is used to drive a $4~kW$, $400~V$ induction motor. The PWM using V/f control is produced using dSpace DS5101.
The experimental setup with the induction motor and the variable dc supply for emulating reconfigurable cells is shown in Fig. \ref{fig:fullsetup}.

The following approach is used to determine the power loss versus dc bus voltage for the SiC inverter.
\begin{enumerate}
    \item A thermal model is developed to quantify the thermal resistance from the case to ambient.
    \item The induction motor is driven at different speeds as per the V/f control algorithm by controlling the dc bus voltage.
    \item Device temperature is measured for each test at a dc bus voltage (and hence motor speed) after ensuring sufficient settling time for the temperature rise.
    \item The thermal model developed in Step 1 is used to compute the total power loss.
    \item The motor rms current is used to determine the conduction losses in the device. 
    \item Switching losses are computed using the total power loss and conduction loss values determined in Steps 4 and 5.
    \item Power losses are compared with the analytical model developed in Section II.
\end{enumerate}

\subsection{Test for thermal model development}
\begin{figure}
    \centering
    \includegraphics[width=0.65\linewidth]{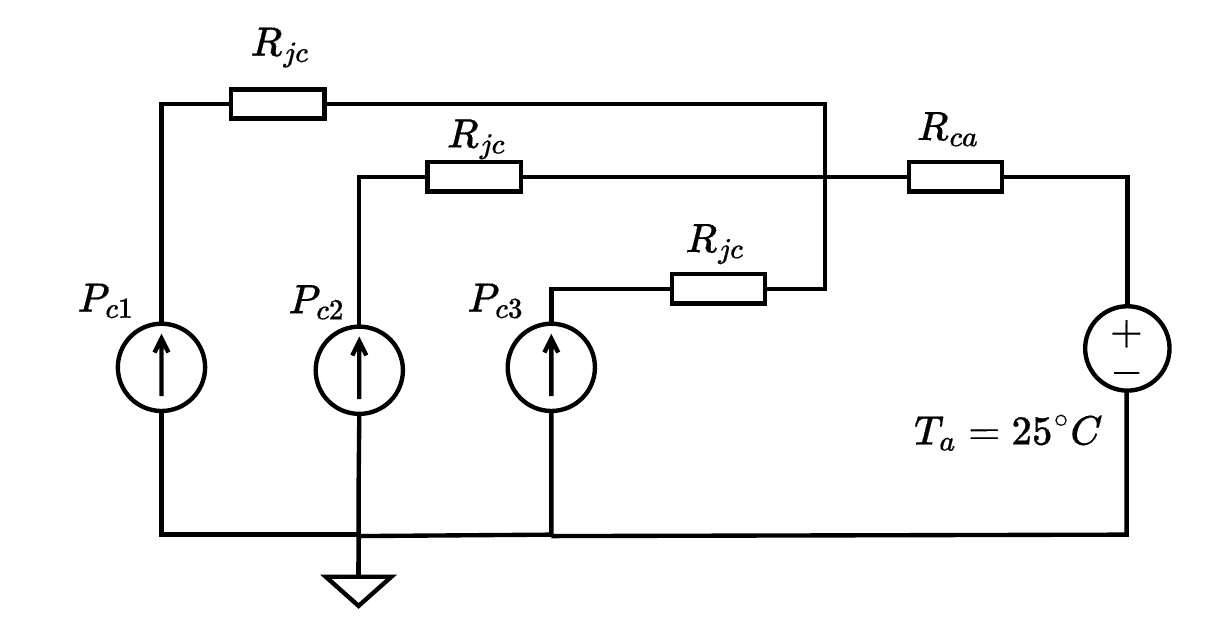}
    \caption{Thermal equivalent circuit for the SiC power module when three devices are turned on with 100\% duty ratio.}
    \label{fig:thermal_eqckt}
\end{figure}

Considering the three phase SiC inverter to be as in Fig. \ref{fig:inv3ph}, the SiC MOSFETs $S_1$, $S_4$ and $S_6$ are turned on at 100\% duty while their complementary devices have 0\% duty. Instead of the motor as load, three-phase resistive load is connected. Since there is no switching involved, only conduction losses cause temperature rise in the SiC module. The dc bus voltage is varied to create different conduction losses. Using the on resistance of the MOSFETs, conduction losses are computed. Thermal equivalent circuit shown in Fig. \ref{fig:thermal_eqckt} is then used to compute the thermal resistance between the case to the ambient ($R_{ca}$). 
Experimental results for two different cases of load resistance are shown in Table \ref{tab:thermal_result}.
\begin{table}[htbp]
    \centering
    \caption{Thermal modeling results for case 1 and case 2.}
    \begin{tabular}{lcc}
        \toprule
        & \textbf{Case 1} &  \textbf{Case 2}  \\
        \midrule
        $I_{a,\mathrm{rms}}$ (A) & 7.79 &  9.70 \\
        $I_{b,\mathrm{rms}}$ (A) & 3.90 & 4.85 \\
        $I_{c,\mathrm{rms}}$ (A) & 3.90  & 4.85 \\
        $R_{\mathrm{ds,on}}$ ($\Omega$) & 0.034  & 0.034 \\
        $P_{\mathrm{cond}}$ (W) & 3.09  & 4.80 \\
        $R_{\mathrm{jc}}$ ($^\circ$C/W) & 0.49  & 0.49 \\
        $T_{\mathrm{case}}$ ($^\circ$C) & 32.4  & 38.6 \\
        $T_{\mathrm{amb}}$ ($^\circ$C) & 22  & 22 \\
        $R_{\mathrm{ca}}$ ($^\circ$C/W) & 3.36  & 3.46 \\
        \bottomrule
    \end{tabular}
    \label{tab:thermal_result}
\end{table}
Note that no forced cooling is used for this test. Based on these tests, average thermal resistance is determined to be
\begin{equation}
   R_{ca}=3.41^\circ C/W \label{eq:rth_ca} 
\end{equation}
Now, by measuring the case temperature, it is possible to compute the power losses in the SiC module by making use of the full thermal equivalent circuit.

\subsection{Results with induction motor drive}
\begin{figure}[htbp]
    \centering
    \begin{subfigure}[b]{\columnwidth}
        \centering
        \includegraphics[width=0.5\columnwidth]{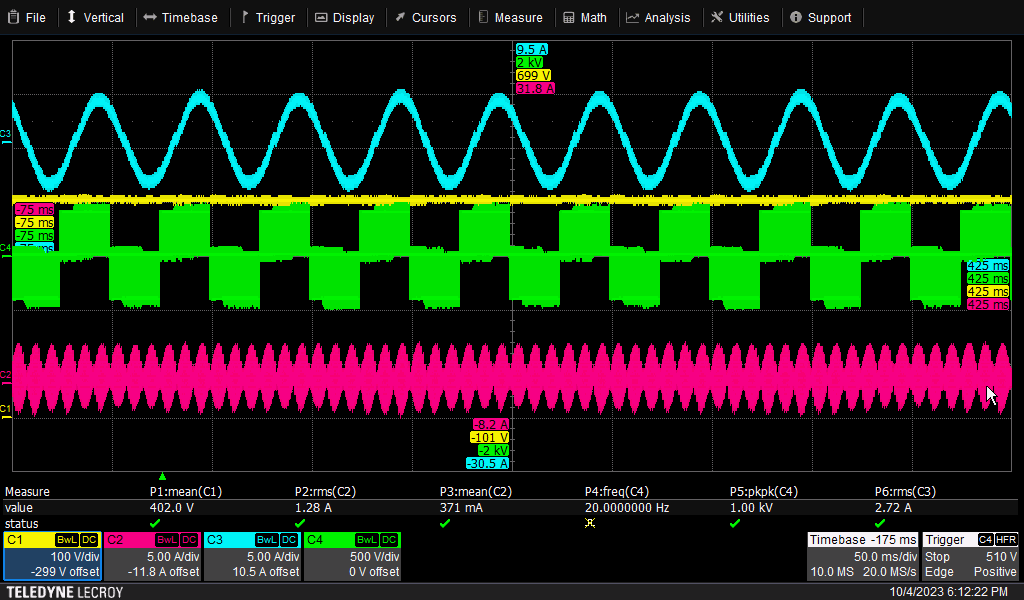}
        \caption{Visualizing fundamental components.}
        \label{fig:exp20hza}
    \end{subfigure}
    
    \begin{subfigure}[b]{\columnwidth}
        \centering
        \includegraphics[width=0.5\columnwidth]{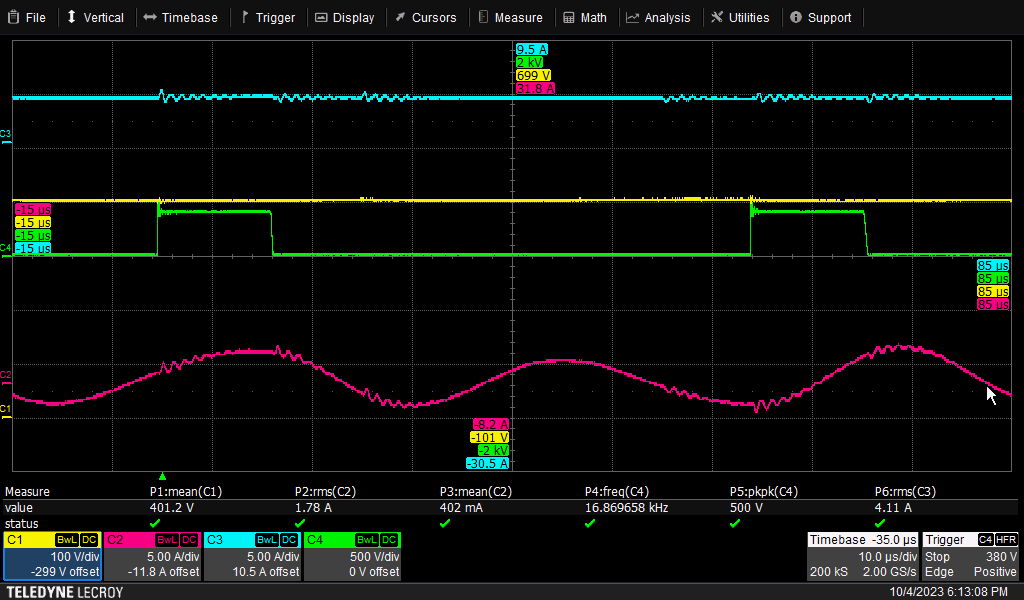}
        \caption{Zoomed to show the switching pulses.}
        \label{fig:exp20hzb}
    \end{subfigure}
    \caption{Experimental results at 20Hz and $V_{dc}=400~V$. C1: DC bus voltage (100V/div), C2: Current from dc supply (5A/div), C3: Motor phase A current (5A/div), C4: Motor line voltage (500V/div).}
    \label{fig:exp20hz}
\end{figure}

\begin{figure}[htbp]
    \centering
    \begin{subfigure}[b]{\columnwidth}
        \centering
        \includegraphics[width=0.5\columnwidth]{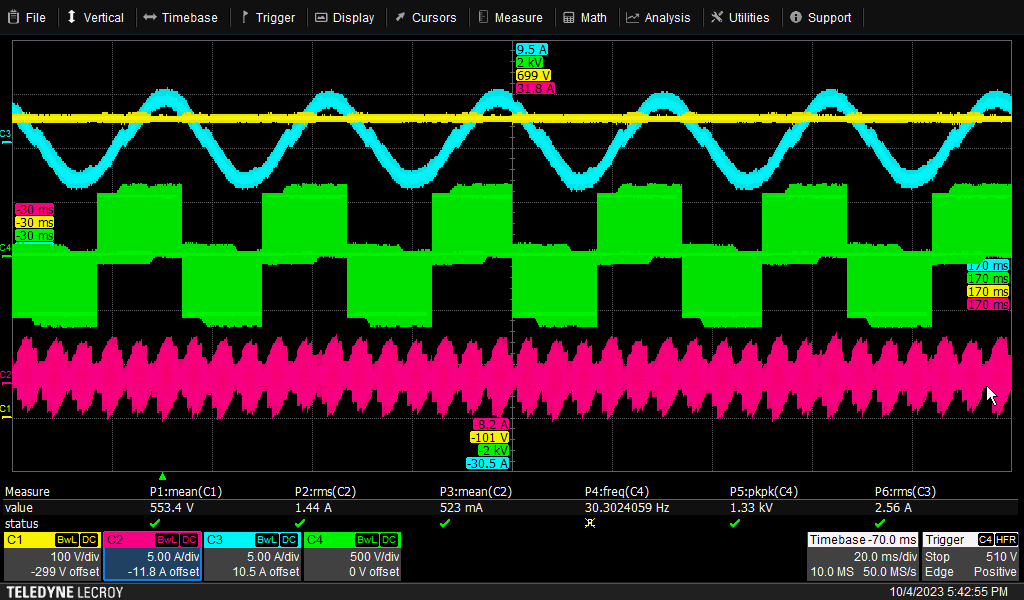}
        \caption{Visualizing fundamental components.}
        \label{fig:exp30hza}
    \end{subfigure}
    
    \begin{subfigure}[b]{\columnwidth}
        \centering
        \includegraphics[width=0.5\columnwidth]{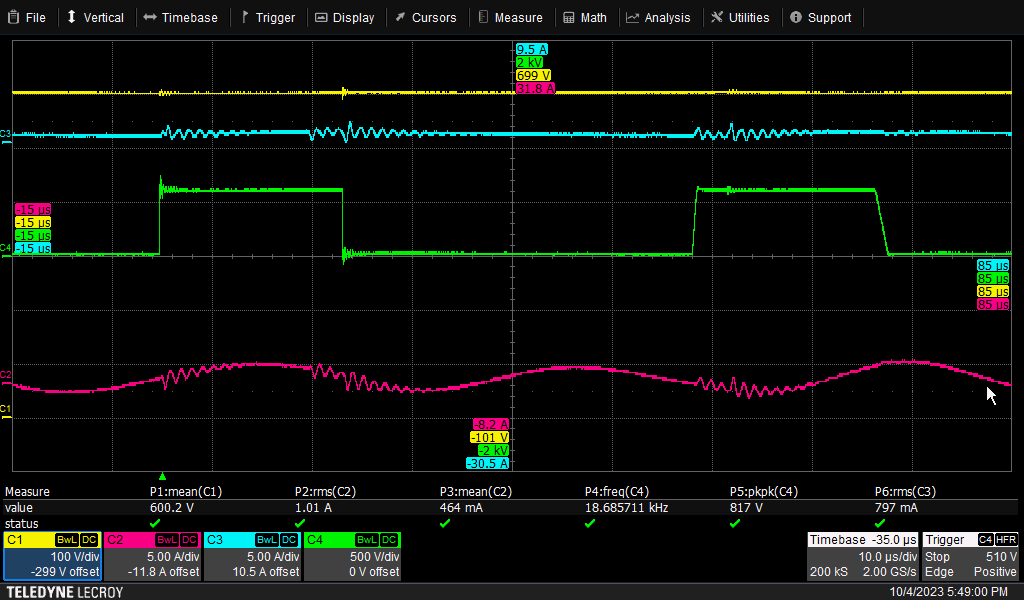}
        \caption{Zoomed to show the switching pulses.}
        \label{fig:exp30hzb}
    \end{subfigure}
    \caption{Experimental results at 30Hz and $V_{dc}=600~V$. C1: DC bus voltage (100V/div), C2: Current from dc supply (5A/div), C3: Motor phase A current (5A/div), C4: Motor line voltage (500V/div).}
    \label{fig:exp30hz}
\end{figure}

%motor time domain waveforms
%power loss waveforms
Fig. \ref{fig:exp20hza} shows the experimental results for the induction motor drive when the frequency is $f=20~Hz$. This corresponds to a motor speed of 600 rpm. The same signals are zoomed in Fig. \ref{fig:exp20hzb} to show the switching pulses at the inverter line-to-line voltage. For this case, the dc bus voltage is set to $V_{dc}=400~V$. 
Similar results are obtained when the induction motor speed is increased. Fig. \ref{fig:exp30hza} shows the experimental results for the induction motor drive when the frequency is $f=30~Hz$. The same signals are zoomed in Fig. \ref{fig:exp30hzb} to show the switching pulses at the inverter line-to-line voltage. For this case, the dc bus voltage is set to $V_{dc}=600~V$. Thus, ideally higher switching losses would be expected in this case due to the higher dc bus voltage across the inverter. 
Table \ref{tab:exp_calc} shows the detailed computations based on the experimental approach described. This table represents the experimental data when the frequency is $20~Hz$ and the dc bus voltage is varied from $200~V$ to $500~V$. First the device temperatures are measured, then total power losses are computed using the thermal model. Finally conduction and switching losses are separated from the total power loss. 
\begin{table*}
\centering
\caption{Experimental data for extracting the device losses at different dc bus voltage levels.}
\begin{footnotesize}
\begin{tabular}{cccccccc}
\toprule
 \textbf{$V_{dc}$ (V)} & \textbf{m} & \textbf{$T_{case}$ (°C)} & \textbf{$P_{loss,tot}$ (W)} & \textbf{$P_{cond,tot}$ (W)} & \textbf{$P_{sw,tot}$ (W)} & \textbf{$P_{sw,mos}$ (W)} & \textbf{$P_{cond,mos}$ (W)} \\
\midrule
 200 & 0.950 & 29.0 & 2.053 & 0.800 & 1.253 & 0.209 & 0.133 \\
250 & 0.760 & 33.5 & 3.372 & 0.800 & 2.573 & 0.429 & 0.133 \\
 300 & 0.633 & 36.4 & 4.223 & 0.800 & 3.423 & 0.571 & 0.133 \\
 350 & 0.543 & 39.0 & 4.986 & 0.800 & 4.186 & 0.698 & 0.133 \\
 400 & 0.475 & 41.6 & 5.748 & 0.800 & 4.948 & 0.825 & 0.133 \\
 450 & 0.422 & 45.2 & 6.804 & 0.800 & 6.004 & 1.001 & 0.133 \\
500 & 0.380 & 48.2 & 7.683 & 0.800 & 6.884 & 1.147 & 0.133 \\
\bottomrule
\end{tabular}
\end{footnotesize}
\label{tab:exp_calc}
\end{table*}
\begin{figure}[htbp]
    \centering
    \begin{subfigure}[b]{0.48\textwidth}
        \centering
        \includegraphics[width=0.9\textwidth]{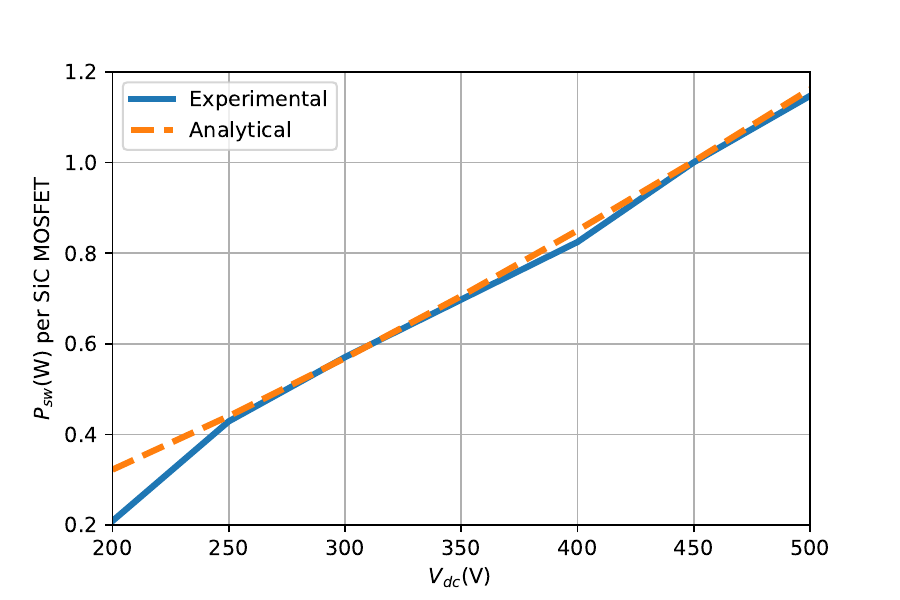}
        \caption{At $f = 20~Hz$.}
        \label{fig:sw_loss_comparisona}
    \end{subfigure}
    %\hfill
    \begin{subfigure}[b]{0.48\textwidth}
        \centering
        \includegraphics[width=0.9\textwidth]{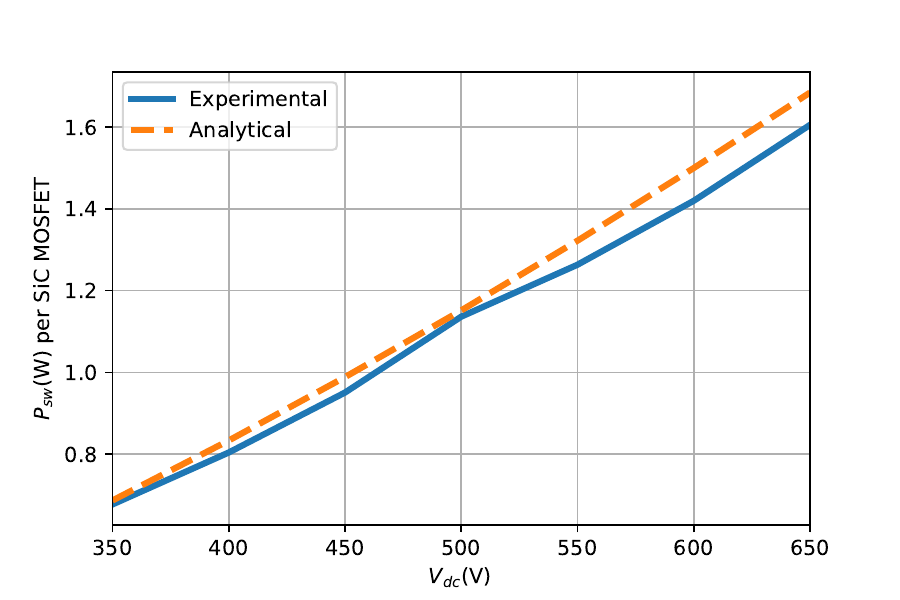}
        \caption{At $f = 30~Hz$.}
        \label{fig:sw_loss_comparisonb}
    \end{subfigure}
    \caption{Comparison of the experimentally determined SiC MOSFET switching losses with the theoretical model.}
    \label{fig:sw_loss_comparison}
\end{figure}
This is repeated at other speeds of the induction motor. The conduction and switching losses for the SiC devices can be computed theoretically using the model developed in Section II. Since only the switching loss depends on the dc bus voltage, its theoretical variation (using the datasheet parameters and (\ref{eq:psw})) is compared with the experimental results. This is shown in Fig. \ref{fig:sw_loss_comparison}.
As it can be seen from Fig. \ref{fig:sw_loss_comparison}, there is a good agreement between the model and the experimental results. Thus the eVTOLs can be designed for a higher dc bus voltage whose value depends on the eVTOL motor ratings and payload capacity. For the example considered in Section II, the optimal dc voltage was determined to be $1~kV$. This voltage will be used during the flight phases of takeoff and landing having the highest power and rotor speed demands. For other operating phases, the dc bus voltage can be reduced using the electronic reconfiguration, thereby improving the power converter efficiency. Thus, the power architecture proposed in this paper is a generic solution and the actual dc bus voltage value will depend on the design of the eVTOL and the motors used.

\section{Conclusion}
In this paper, an optimization of the eVTOL power architecture   is proposed, focusing on the impact of dc bus voltage on overall system efficiency and cable weight. An accurate analytical model was developed to formulate an objective function and perform the optimization. By systematically analyzing the performance of SiC-based inverters within a varied dc bus voltage range, an optimal dc bus voltage was identified that balances inverter efficiency and cable weight reduction. For the eVTOL specifications considered in the paper, the optimal dc bus voltage is evaluated as $1~kV$. The proposed power converter topology and reconfigurable battery architecture demonstrate improvements in both efficiency and safety. The reconfigurable dc bus architecture allows for dynamic adjustment of the dc bus voltage according to different phases of the eVTOL mission profile, hence improving battery utilization.
For the eVTOL parameters considered in this paper, the proposed approach reduces the overall weight of the power cables by up to 2.5 and also ensures safer and more reliable operation by closely integrating the  BMS with the FCC. Proposed method is validated experimentally using a scaled-down laboratory setup. 
Proposed approach can be used for designing more efficient and lightweight power architectures for eVTOLs, which are crucial for their commercial viability and operational efficiency.

%%%%%%References%%%%%%%%%%%%%%%%%%%%%%%%%%%%%%%%%%%%%%%%%%%%%%%%%%
%%\bibliographystyle{IEEEtran} % Tell LaTeX to use the IEEEtran style
%%\bibliography{references.bib}

\end{document}